\documentclass[12pt]{iopart}
\usepackage{graphicx}
\usepackage{amsfonts}
\usepackage{color}

\begin{document}

\title{Entanglement versus Quantum Discord in Two Coupled Double Quantum Dots}
\author{F. F. Fanchini$^1$, L. K. Castelano$^2$, and
A. O. Caldeira$^1$}
\address{$^1$ Instituto de F\'isica Gleb Wataghin, Universidade
Estadual de Campinas, P.O. Box 6165, CEP 13083-970, Campinas, SP,
Brazil}
\address{$^2$ Department of Physics, University of
California-San Diego, La Jolla, California 92093-0319, USA}

\begin{abstract}
We study the dynamics of quantum correlations of two coupled
double quantum dots containing two excess electrons. The
dissipation is included through the contact with an
oscillator bath. We solve the Redfield master
equation in order to determine the dynamics of the quantum discord
and the entanglement of formation. Based on our results, we find
that the quantum discord is more resistant to dissipation than the
entanglement of formation for such a system. We
observe that this characteristic is related to whether the
oscillator bath is common to both qubits or not
and to the form of the interaction Hamiltonian.
Moreover, our results show that the quantum discord might be
finite even for higher temperatures in the asymptotic limit.
\end{abstract}
 \DeclareGraphicsExtensions{.jpg, .pdf, .mps, .png, .tiff}
 \maketitle

\section{Introduction}
Among the ``unusual'' manifestations  observed in the quantum
world,  entanglement is undoubtedly one of the most interesting
ones. Entanglement corresponds to global states of two or more
quantum systems that cannot be separated into direct product
states of individual subsystems. This characteristic yields
correlations between quantum systems that cannot be found in any
classical system \cite{brooks}. Moreover, entanglement is a very
important ingredient for quantum computer's architecture
\cite{horodecki} and quantum communication \cite{brooks}. It is
well-known that some tasks executed by a quantum computer can
be performed exponentially faster than by the existing computers.
These potential applications of a quantum computer have motivated a large number
of experimentalists and different approaches have been used to
build entangled states in laboratories \cite{experiments}. Despite
the successful experimental achievements, there are many
difficulties to overcome before a functional quantum computer
becomes a reality. One of the most trivial difficulties is the
fact that the usual quantum systems candidates for qubits (see
below) are intrinsically open to their environments and therefore
suffer their effects by losing coherence. Thus, pure quantum
states become mixed states under the environment's influence.
Fortunately, it has been demonstrated that computers based on
mixed states can also be used to solve certain tasks more
efficiently than classical computers \cite{knill}, although they
are less powerful than the computation using pure states. The reason for such a
performance is attributed to correlations not presented in
classical systems. These quantum correlations (QCs) can be identified
through a quantity called quantum discord \cite{zurek}. Recently,
the interest on this subject has received great attention due to
the possibility of achieving quantum computation without entangled
states \cite{lanyon}. As a consequence, many different aspects of
quantum discord have been discussed, \emph{e.g.}, robustness to
quantum sudden death \cite{fff,ferraro}, relation to the speed-up in
deterministic quantum computation with one qubit \cite{animesh},
the interplay between quantum phase transitions and quantum
discord \cite{sarandy}, and the condition to obtain completely
positive maps \cite{shabani}.

\begin{figure}[t]
\begin{center}
\includegraphics[width=8cm]{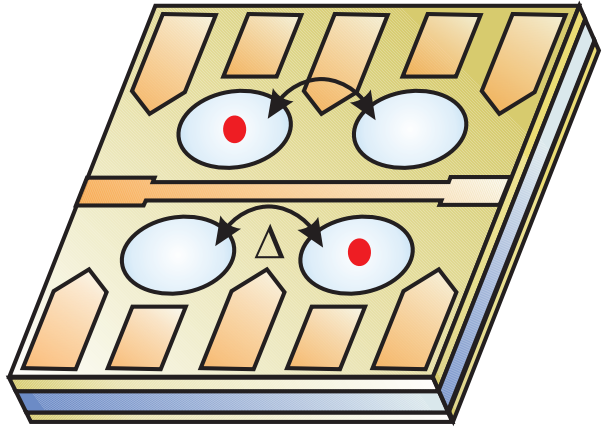}
\caption{(Color online) Schematic representation
of two DQDs with one electron localized at the left (right) side of the top (bottom) DQD.
 See Ref.~\cite{shinkai} for further details.}\label{fig0}
 \end{center}
\end{figure}

There are many different systems proposed as candidates for
qubits. Among them, double quantum dots (DQDs) (see Fig.~\ref{fig0})
 are very interesting due to the easy integration with
existing electronics and the advantage of scalability
\cite{itakura}. Moreover, arrays of quantum dots can be employed
to perform logical operations, which can be used to implement an
universal quantum computation based on quantum \cite{toth} and classical effects
\cite{lent}. In this work, we investigate the quantum correlation
dynamics of two DQDs including the basic elements to simulate a
realistic situation, \emph{e.g.}, finite temperatures, interaction
between qubits, and detuning.  We also compare the effects of dissipation on the behavior of
the quantum discord (QD) with that of the entanglement of formation
(EoF) \cite{eof}. Our results show the longer duration of QD
against the EoF. To explain such a result, we analyze how the
environment acts on the system by comparing the case where the
bath is coupled to both DQDs, which is the more realistic case, to
the case where each DQD is coupled to its own bath. Based on this
comparison, we verify that the common character of the bath might
be very important to prolong the QD. In fact, this character of
the environment brings different properties to the dissipative
dynamics. Contrary to the case of independent environments, by
which the coherences are completely lost, we can find regions of
the state space where the coherences are preserved \cite{dfs}. We
observe that this robustness can be transferred to the quantum
discord.

The present paper is organized as follows. In Sect.~II we describe
the basic concepts of quantum discord. In Sect.~III we present the
Hamiltonian of the system and the master equation which accounts
for the interaction between the system and the environment. In
Sect.~IV, we discuss the obtained results for the quantum
correlations of two coupled DQDs. Finally, we summarize our work
in Sect.~V.

\section{Quantum Discord}
Quantum discord has been proposed as a quantity that captures the
quantum correlations between two subsystems \cite{zurek}. To
evaluate such QCs, one needs to subtract the
classical correlations from the total correlation
\cite{zurek,vedral}. The total correlation of a bipartite quantum
systems, $A$ and $B$, is calculated by the quantum mutual
information \cite{mutual}:
\begin{equation}
\mathcal{I}(A:B) = S(A) - S(A|B),
\end{equation}
where  $S(A|B)= S(AB) - S(B)$ and $S(X)=-{\rm Tr}\{X\log X\}$ is
the von Neumann entropy of the density matrix $X$. Here, we adopt
the following definitions: $AB\equiv\rho_{AB}$,  $A \equiv
\rho_A={\rm Tr_B}\{\rho_{AB}\}$, and $B \equiv \rho_B={\rm
Tr_A}\{\rho_{AB}\}$, where $\rho_{AB}$ is the bipartite density
matrix.

To calculate the classical correlations we observe  that the
projective measurements on a subsystem remove all nonclassical
correlations between the parts, \emph{i.e.}, after a measurement on a particular subsystem, all QCs are destroyed. Thus, we define a quantity that
evaluates the mutual information after a measurement on one of the
subsystems
\begin{equation}
\mathcal{I}(A:B)_{\{\Pi^B_j\}} = S(A) - S(A|\{\Pi_j^B\}),
\end{equation}
where $S(A|\{\Pi_j^B\})$ is the system conditional entropy after
the measurement and $\{\Pi_j^B\}$ defines a complete set of one-dimensional
projectors, and the different outcomes of this measurement are
accounted for by $j$. To quantify the quantum correlations, since
$\mathcal{I}(A:B)_{\{\Pi^B_j\}}$ depends on the projector basis
$\{\Pi_j^B\}$, we take the maximum of
$\mathcal{I}(A:B)_{\{\Pi^B_j\}}$ taking into account all possible
projectors. Thus, we can define the following quantity,
\begin{equation}
\mathcal{J}(A:B) = S(A) - {\rm min}_{\{\Pi_j^B\}}S(A|\{\Pi_j^B\}),
\end{equation}
that gives a measure of the total classical correlations between
two subsystems \cite{vedral}.

Therefore, the QD can be written as:
\begin{equation}
\delta(A:B)=\mathcal{I}(A:B) - \mathcal{J}(A:B).\label{discord}
\end{equation}
It is straightforward to observe that if all information can be
obtained locally by $B$, this subsystem has QD equals to zero.
This implies that a measurement on $B$ does not alter the state of
$A$. Otherwise, if just part of the information can be obtained
locally, $B$ is quantum correlated to $A$. The QD thus gives the
total information that is not locally accessible and has been
accepted as a measure of the quantum correlation. Besides, the QD
is possibly finite even for separable quantum states and can be
utilized as a new resource for quantum computation \cite{knill,
lanyon}. Also, when only pure states are considered, QD and
the entanglement of formation \cite{eof} are indistinguishable.

\section{Theoretical Model}
\subsection{Hamiltonian of the System}
The system considered in this work is composed of two double
quantum dots (see Fig.~1), where each DQD has an excess electron
localized in either the left  $|L\rangle$ or the right dot
$|R\rangle$. This system can be modelled by the following
pseudospin Hamiltonian
\begin{equation}
H_S=\Delta\left(\sigma_{x}^{(1)}+\sigma_{x}^{(2)}\right)+
J\sigma_{z}^{(1)}\otimes\sigma_{z}^{(2)}\label{h0},
\end{equation}
where the first term describes the tunnelling coupling energy
$\Delta$ between the qubits, and the last term takes into account
the effects of the Coulomb interaction between the two electrons,
which favors anti-parallel configurations $|L,R\rangle$ and
$|R,L\rangle$ over the parallel ones $|L,L\rangle$ and
$|R,R\rangle$. We adopt the convention $|L\rangle\equiv
|\downarrow\rangle$ and $|R\rangle\equiv |\uparrow\rangle$. For
simplicity, we set the energy offset of the single-qubit states
equals to zero.

\subsection{Master Equation}
In order to analyze the two DQDs dynamics in an open quantum
system, we suppose that both qubits are coupled to a bath of
harmonic oscillators (phonons). The total Hamiltonian that
computes the environment perturbation is given by
\begin{eqnarray}
H=H_S+\left(\sigma_z^{(1)}+\sigma_z^{(2)}\right)\mathcal{L} + H_B,\label{htot}
\end{eqnarray}
where $\mathcal{L}=B+B^\dagger$ with $B=\sum_kg_ka_k$,
$H_B=\sum_k\omega_k a_k^\dagger a_k$, $\omega _{k}$ is the
frequency of the $k^{th}$ normal mode of the
bath , and $\hbar=1$.  Here $a_k$ is the usual
annihilation operator of this mode and $g_k$ is
its complex coupling constant with the DQDs. It is important to
note that the interaction Hamiltonian acts in a correlated way
with both qubits, thereby representing a common bath. To
numerically determine the evolution of the two DQDs reduced
density matrix, we employ the Redfield master equation
\cite{shibata77}
\begin{eqnarray}
\frac{d\rho _{I}(t)}{dt}=-\int^{t}_{0}dt^{\prime} {\rm
Tr}_{B}\left\{{\left[H_{I}(t),\left[H_{I}(t^{\prime}),\rho
_{B}\rho _{I}(t)\right]\right]}\right\},\label{master}
\end{eqnarray}
where $H_{I}(t)$ is the interaction Hamiltonian in the interaction picture,
\begin{eqnarray}
H_{I}(t)=U_S^{\dagger}(t)U^{\dagger}_{B}(t)\left\{
\left(\sigma_z^{(1)}+\sigma_z^{(2)}\right)\mathcal{L}\right\}U_{B}(t)U_S(t),\label{hintint}
\end{eqnarray}
where $U_{B}(t)=\exp\left(-iH_{B}t\right)$ and $U_S(t)=\exp\left(-iH_{S}t\right)$.
Explicitly, the unitary evolution $U_S(t)$ is given by
\begin{eqnarray}
U_{S}(t) & = & \frac{1}{2}\left[\cos\left(\Omega t\right)+\cos\left(Jt\right)\right]{\rm I}\nonumber\\
 & - & i\frac{\Delta}{\Omega}\left[\sin\left(\Omega t\right)\right]
 \left(\sigma_{x}^{(1)}+\sigma_{x}^{(2)}\right)\nonumber\\
 & + & \frac{1}{2}\left[\cos\left(\Omega t\right)-\cos\left(Jt\right)\right]\sigma_{x}^{(1)}\sigma_{x}^{(2)}\nonumber\\
 & - & \frac{i}{2}\left[\sin\left(Jt\right)-\frac{J}{\Omega}\sin
 \left(\Omega t\right)\right]\sigma_{y}^{(1)}\sigma_{y}^{(2)}\nonumber\\
 & - & \frac{i}{2}\left[\sin\left(Jt\right)+\frac{J}{\Omega}\sin
 \left(\Omega t\right)\right]\sigma_{z}^{(1)}\sigma_{z}^{(2)},
\end{eqnarray}
where $\Omega=\sqrt{J^{2}+4\Delta^{2}}$. Here, we suppose that the
oscillator bath density matrix $\rho _{B}$ is
initially decoupled from the system,
\begin{eqnarray}
\rho_{B}=\frac{1}{Z}\exp(-\beta H_{B}),\label{rhoB}
\end{eqnarray}
where $Z$ is the partition function $Z={\rm Tr}_{B}
\left[\exp(-\beta H_{B})\right]$, $\beta =1/k_{B}T$, $k_{B}$ is
Boltzmann constant, and $T$ is the absolute temperature of the
environment. Defining $U_{B}^\dagger(t){\mathcal
L}U_{B}(t)\equiv\tilde{\mathcal L}(t)$ and $\Lambda(t)\equiv
U_S^{\dagger}(t)
\left(\sigma_z^{(1)}+\sigma_z^{(2)}\right)U_S(t)$, we can write
the interaction Hamiltonian in the interaction picture as follows:
\begin{equation}
H_{I}(t)=\Lambda(t)\tilde{\mathcal L}(t).\label{hint}
\end{equation}
Thus, substituting Eq.~(\ref{rhoB}) and Eq.~(\ref{hint}) in the master equation, Eq.~(\ref{master}), we obtain
\begin{eqnarray}
\frac{d\rho _{I}(t)}{dt}&=&\int^{t}_{0}dt^{\prime} \mathcal{D}(t,t^\prime)
\left[\Lambda(t),\rho _{I}(t)\Lambda(t^\prime)\right]\nonumber\\
&+&\int^{t}_{0}dt^{\prime}
\mathcal{D}^\ast(t,t^\prime)\left[\Lambda(t^\prime)\rho
_{I}(t),\Lambda(t)\right],
\end{eqnarray}
where $\mathcal{D}(t,t^\prime)=T_1(t-t^\prime)+T_2(t-t^\prime)$ with
\begin{eqnarray}
T_1(t-t^\prime)&=&{\rm Tr}_{B}\!\!\left\{ \tilde{B}(t){\tilde\rho} _{B}
{\tilde B}^\dagger(t^{\prime})\!\right\}\nonumber\\
&=&\sum_k|g_k|^2n_k\exp[-i\omega_k(t-t^\prime)],\label{t1}
\end{eqnarray}
and
\begin{eqnarray}
T_2(t-t^\prime)&=&{\rm Tr} _{B}\!\!\left\{ \tilde{B}^\dagger(t){\tilde\rho}_{B}
{\tilde B}(t^{\prime})\!\right\}\nonumber\\
&=&\sum_k|g_k|^2(n_k+1)\exp[i\omega_k(t-t^\prime)],\label{t2}
\end{eqnarray}
with $\tilde{B}(t)=U_{B}^\dagger(t)B U_{B}(t)$, ${\tilde
\rho}_B=U_{B}^\dagger(t)\rho_B  U_{B}(t)$, and $n_k$ is the
average occupation number of mode $k$:
\begin{equation}
n_k=\frac{1}{\exp(\beta\omega_k)-1}.
\end{equation}
Defining, as usual, the spectral function
\begin{eqnarray}
J(\omega)\equiv\sum_{k}|g_{k}|^{2}\delta(\omega-\omega_{k}),
\end{eqnarray}
we can replace the summations above by integrals
\begin{eqnarray}
T_1(t)=\int_0^\infty d\omega J(\omega)n(\omega)\exp(-i\omega t),
\end{eqnarray}
and
\begin{eqnarray}
T_2(t)=\int_0^\infty d\omega J(\omega)\exp(i\omega
t)[n(\omega)+1].
\end{eqnarray}
If we assume an ohmic spectral density  to the reservoir,
$J(\omega)=\eta\omega\exp(-\omega/\omega_c)$, where $\omega_c$ is
a cutoff frequency and $\eta$ is the damping constant, we can
explicitly evaluate the integrals above which yield:
\begin{eqnarray}
\hspace{-0.6cm}\mathcal{D}(t,t^\prime)\!&=&\frac{\eta\omega_c^2}{[1+i\omega_c(t-t^\prime)]^2}\nonumber\\
&+&\!\frac{2\eta}{\beta^2}{\rm Re}\left\{\Psi^{(1{\rm st})}(1+1/(\beta\omega_c)-i(t-t^\prime)/\beta)\right\}\!,
\end{eqnarray}
where $\Psi^{(1{\rm st})}(\cdot)$ is the first polygamma function.

\begin{figure}[t]
\begin{center}
\includegraphics[width=15cm]{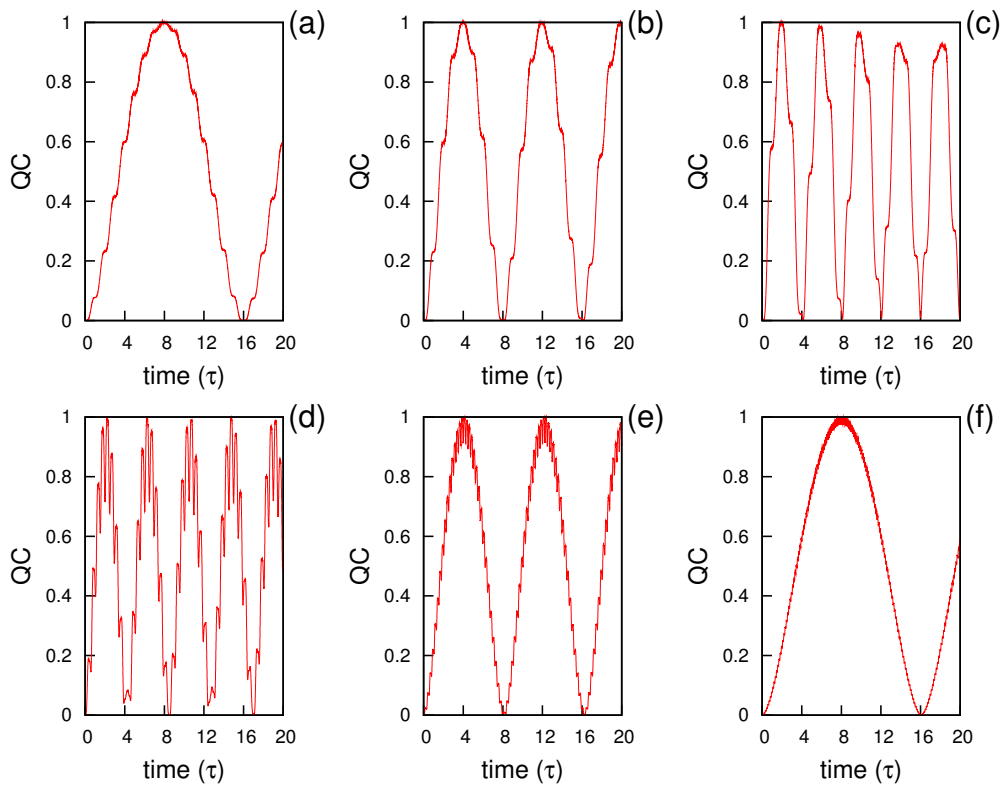}
\caption{(Color online) Quantum correlations (QD and EoF) for the
unitary evolution ruled by $H_S$. We consider different values of
$J/\Delta$ and we fix the coupling constant $\eta=0$. In (a), (b), and (c) we have $J=\Delta/8$,
$J=\Delta/4$, and $J=\Delta/2$ respectively. This is what we
define as weak coupling regime. For the strong coupling regime we
have (d), (e), and (f) with $J=4\Delta$, $J=8\Delta$, and
$J=16\Delta$ respectively.} \label{fig1}
\end{center}
\end{figure}

\section{Quantum Correlation Dynamics of DQDs}
To calculate the DQDs quantum correlation dynamics we numerically
solve Eq.~(\ref{discord}) using the density matrix dynamics
computed by Eq.~(\ref{master}) as input. In our simulations we
fixed $\Delta=\pi\hbar/2\tau$ ($\Delta\approx$ 10$\mu$eV) and the cutoff frequency $\omega_c=200/\tau$ \cite{contreras}, where
$\tau=10^{-10}s$. These parameters correspond to
typical experimental values in GaAs/AlGaAs lateral DQDs
\cite{shinkai}. Our expressions are given as a function of the
weight of the Coulomb interaction between the two electrons, $J$,
and the damping constant $\eta$. First, we analyze the unitary
dynamics in Fig.~\ref{fig1} supposing that the two DQDs are
initially uncorrelated,
$\rho_I(0)=|\!\uparrow\downarrow\rangle\langle\uparrow\downarrow\!|$,
and assuming that the coupling with the environment is zero,
\emph{i.e.}, $\eta=0$. Since we suppose an initial pure state, the
EoF and the QD dynamics coincide.
In Fig.~\ref{fig1} we observe that similar dynamics can be reached for different Coulomb coupling values $J$.  Thus,
one can find a entanglement dynamics in the weak coupling regime ($J<<\Delta$) that corresponds to a similar dynamics in the strong-coupling regime ($J>>\Delta$), and vice versa.
To determine these equivalent entanglement dynamics, we examine the period that the initial
state $|\!\uparrow\downarrow\rangle$ reaches a maximum entangled
state $t_{max}$ for $J<<\Delta$ and $J>>\Delta$. For this purpose, we estimate
 these periods analyzing the coherences of
$U_S(t)\rho(0)U_S^\dagger(t)$. In the weak coupling regime $J<<\Delta$ we obtain
\begin{equation}
t_{max}\approx\frac{1}{J_{<}}\left( \frac{\pi}{2}+n\pi\right)\label{weakregime},
\end{equation}
where $n \in \mathbb{N}$ and $J_{<}$ denotes that $J<<\Delta$.
For the strong coupling regime $J>>\Delta$ ($J_{>}$), the period that the initial state reaches a maximum entangled state depends on $\Delta$ and is given by
\begin{equation}
t_{max}\approx \frac{J_{>}}{\Delta^2}\left(\frac{\pi}{4} + n\frac{\pi}{2}\right)\label{strongregime}.
\end{equation}
Based on Eqs.~(\ref{weakregime}) and (\ref{strongregime}) we find that the time to achieve a maximum entangled state is approximately the same in both regimes when $J_{>}J_{<}=2\Delta^2$.
 This behavior can be noted by comparing Fig.~\ref{fig1}-a ($J_{<}=\Delta/8$) with Fig.~\ref{fig1}-f ($J_{>}=16\Delta$) and Fig.~\ref{fig1}-b ($J_{<}=\Delta/4$) with Fig.~\ref{fig1}-c ($J_{>}=8\Delta$).
\begin{figure}[t]
\begin{center}
\includegraphics[width=13cm]{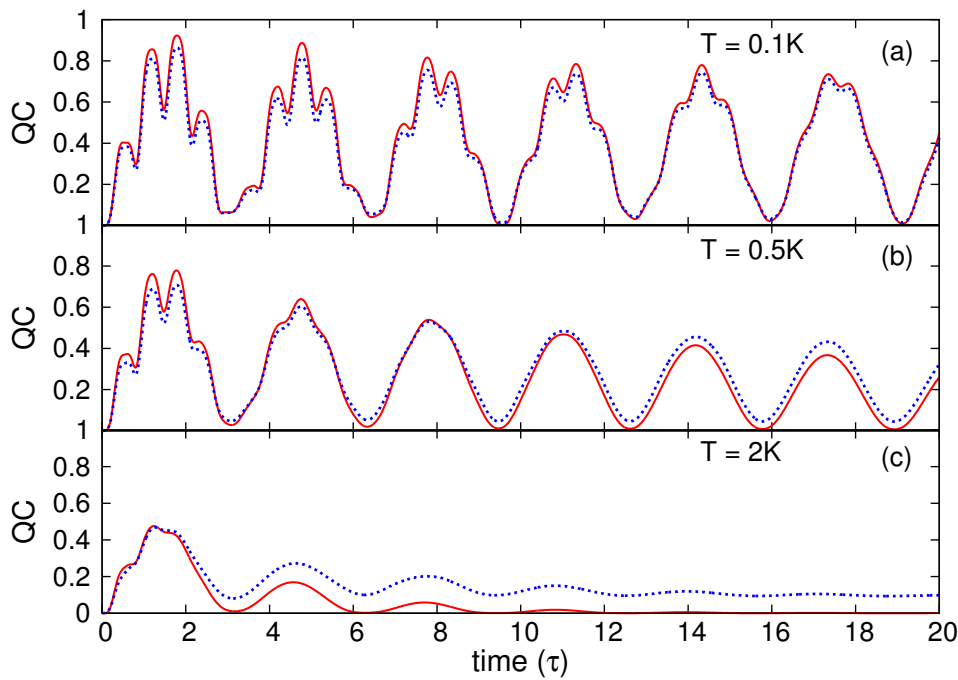}
\caption{(Color online) Two-qubit quantum correlations (QD and
EoF) under a dissipative dynamics for $J=4\Delta$ and $T=0.1$K in
(a), $T=0.5$K in (b), and $T=2$K in (c). The blue (dotted) is the QD dynamics and
the red (solid) line is the EoF dynamics. Here we use $\eta=1/600$.}\label{fig2}
\end{center}
\end{figure}

\begin{figure}[t]
\begin{center}
\includegraphics[width=13cm]{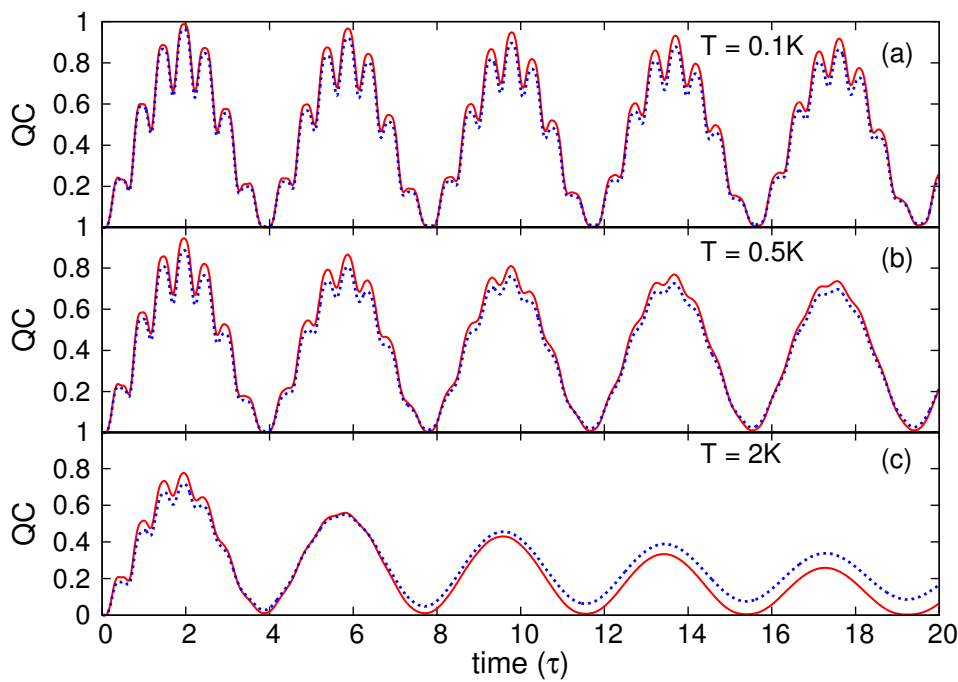}
\caption{(Color online) Two-qubit quantum correlations (QD and
EoF) under  a dissipative dynamics for $J=4\Delta$ and $T=0.1$K in
(a), $T=0.5$K in (b), and $T=2$K in (c). The blue (dotted) is the QD dynamics and
the red (solid) line is the EoF dynamics. Here we use $\eta=1/200$.}\label{fig21}
\end{center}
\end{figure}

\begin{figure}[t]
\begin{center}
\includegraphics[scale=.92]{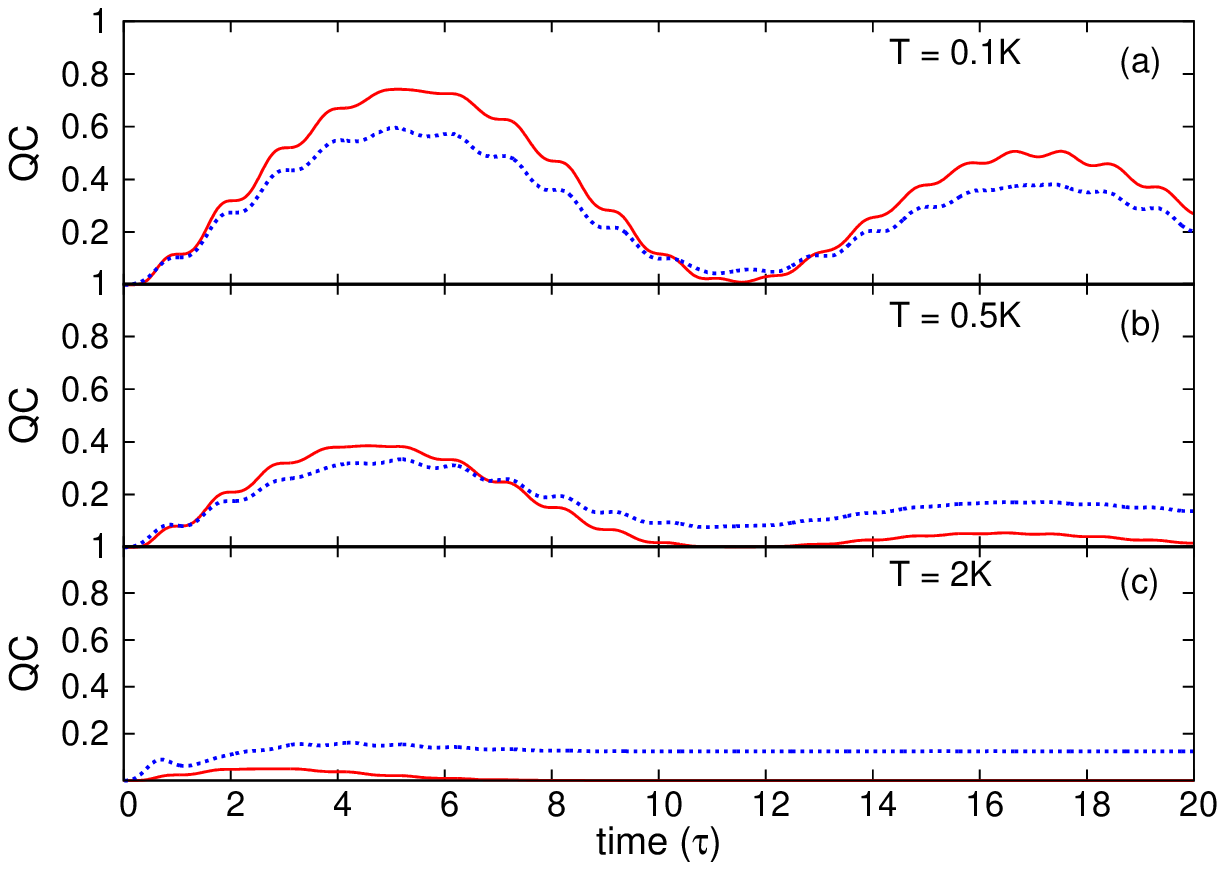}
\caption{(Color online) Two-qubit quantum correlations (QD and
EoF) under a  dissipative dynamics for $J=\Delta/4$ and $T=0.1$ K in
(a), $T=0.5$ K in (b), and $T=2$ K in (c). The blue (dotted) is the QD dynamics and
the red (solid) line is the EoF dynamics. Here we use $\eta=1/600$.}\label{fig3}
\end{center}
\end{figure}

\begin{figure}[b]
\begin{center}
\includegraphics[scale=0.92]{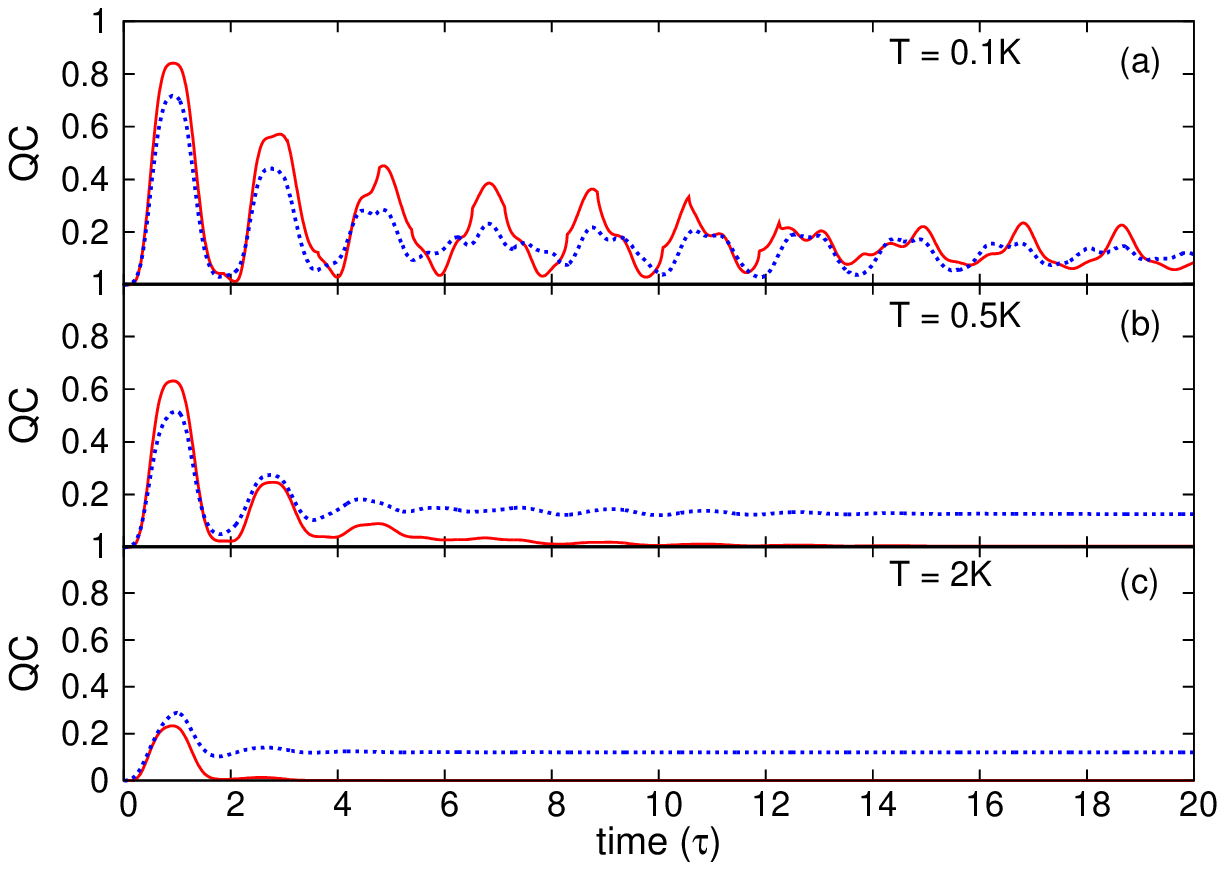}
\caption{(Color online) Two-qubit quantum correlations (QD and
EoF) under a  dissipative dynamics for $J=\Delta/4$ and $T=0.1$ K in
(a), $T=0.5$ K in (b), and $T=2$ K in (c). The blue (dotted) is the QD dynamics and
the red (solid) line is the EoF dynamics. Here we use $\eta=1/200$}\label{fig31}
\end{center}
\end{figure}

In Fig.~\ref{fig2}, assuming that
$\rho_I(0)=|\!\uparrow\downarrow\rangle\langle\uparrow\downarrow\!|$,
we present the numerical results for the QD and the EoF dynamics
within the strong coupling regime ($J=4\Delta$), for different
temperatures ($T=0.1$K, $T=0.5$K, and $T=2$K) and for finite
damping constant ($\eta=1/600$). The EoF for two qubits can be
expressed in terms of the concurrence \cite{conc}, ${\rm{EoF}} =
-f(C)\log_2(C)-(1-f(C))\log_2(1-f(C))$, where
$f(C)=(1+\sqrt{1-C^2})/2$. On the other hand, the concurrence $C$
is defined as the maximum between zero and
$\lambda_1-\lambda_2-\lambda_3-\lambda_4$, where $\lambda_i$, for
$i=1,2,3,4$, is the square root of the eigenvalues of the matrix
$\rho\sigma_y\otimes\sigma_y\rho^\ast\sigma_y\otimes\sigma_y$,
with $\lambda_1$ being the largest one among them, and $\rho^\ast$
is the complex conjugate of $\rho$. In Fig.~\ref{fig2} one notices
that the QCs (QD and EoF) oscillate rapidly, which resembles the
behavior observed when there is no coupling to the environment
(see Fig.~\ref{fig1} (d)). However, when the coupling to the bath
is included, the QCs decay as a function of time due to the loss
of correlations between the qubits and the environment. Moreover,
we verify that QD is more robust than EoF for higher temperatures
and for higher values of the coupling to the environment $\eta$
(see Fig.~\ref{fig21}), and remarkably reaches a constant value in
the asymptotic limit, when $t\rightarrow\infty$. In
Fig.~\ref{fig3} and Fig.~\ref{fig31}, the EoF and QD dynamics are
analyzed for different temperatures and bath couplings $\eta$
within the weak regime $J=\Delta/4$. The QCs for $\eta=1/600$
presented in Fig.~\ref{fig3} have a
 oscillation period that corresponds to the unitary evolution (Fig.~2(b)). However, when the coupling to the environment is more relevant $\eta=1/200$, we observe a short oscillation period (see Fig.~\ref{fig31}).
 Again the QCs decay as a function of time due to the interaction with the environment in both Figs.~\ref{fig3} and \ref{fig31}.
By comparing the results of the weak coupling regime to those in the strong coupling regime, we notice that the QD is also more robust than EoF for higher temperatures and the QD tends to a finite value in this case.
This behavior is related to the convergence of the system when $t\rightarrow\infty$
 to quantum subspaces defined by the eigenvectors of $\sigma_z^{(1)} + \sigma_z^{(2)}$, {\it i.e.}, the system operator that is coupled to the environment \cite{zurekein}. In such a case, the resultant system density matrix can be written as
\begin{equation}
\rho(t\rightarrow\infty) \approx p_1|\!\uparrow\uparrow\rangle\langle\uparrow\uparrow\!| + p_2|\Psi_\pm\rangle \langle\Psi_\pm| + p_3|\!\downarrow\downarrow\rangle\langle\downarrow\downarrow\!|,
\end{equation}

\begin{figure}[t]
\begin{center}
\includegraphics[width=12cm]{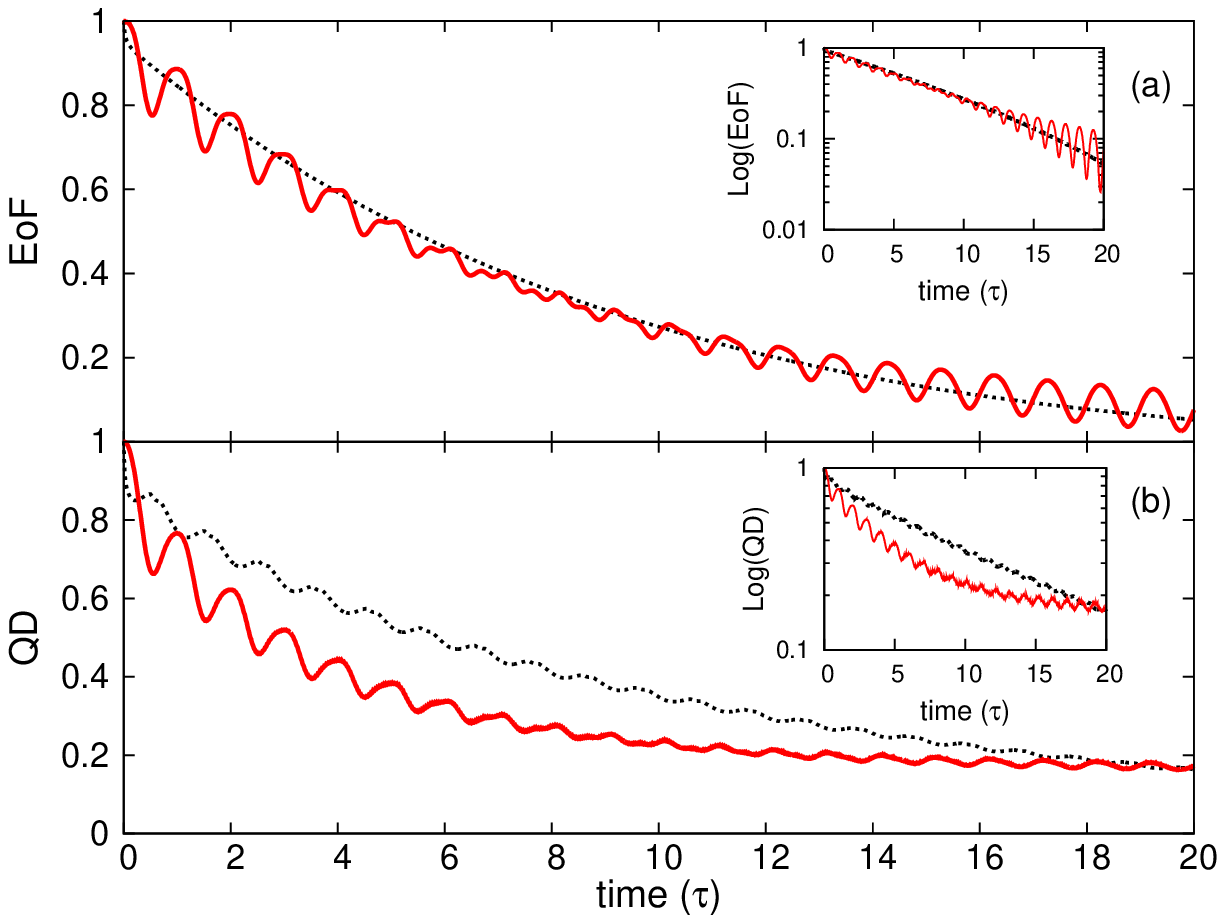}
\caption{(Color online) EoF (a) and QD (b) as a function of time,
considering $J=0$ for independent (dotted curve) and common (solid
curve) baths with $T=0.1$ K and $\eta=1/600$. The insets show the results in logarithmic scale.}\label{fig4}
\end{center}
\end{figure}

where $|\Psi_\pm\rangle=\alpha|\uparrow\downarrow\rangle +\beta|\!\downarrow\uparrow \rangle$, and $p_1$, $p_2$, and $p_3$ are the weights of each eigenvector in the asymptotic limit. Because $\sigma_z^{(1)} + \sigma_z^{(2)}$ is degenerated, the density matrix coherence, whose modulus is given by $p_2|\alpha\beta|$, is sustained. This fact induces a finite quantum correlation that is preserved by the two DQDs. On the other hand, the equilibrium density matrix describes a disentangled state since $p_2|\alpha\beta|\leq\sqrt{p_1p_3}$ \cite{yu}.
To verify that this is the cause of the robustness of the
two DQDs quantum discord, we compare two distinct situations. In
the first situation, each DQD is coupled to its own heat bath,
which are completely uncorrelated. The second situation has a
common environment,
\emph{i.e.}, the same bath is coupled to both DQDs and the interaction Hamiltonian is the one already considered $\left(\sigma_z^{(1)}+\sigma_z^{(2)}\right)\mathcal{L}$.

For independent environments, the interaction
Hamiltonian in Eq.~(\ref{htot}) is slightly different and is given
by $\sigma_z^{(1)} \mathcal{L}^{(1)} + \sigma_z^{(2)}
\mathcal{L}^{(2)}$. Therefore, the master equation's correlation
functions, Eq.~(\ref{t1}) and Eq.~(\ref{t2}), are written as
$T_1(t-t^\prime)={\rm Tr}_{B}\!\!\left\{\sum_{k,k^\prime=1}^{2}
{B}^{(k)}(t){\rho}_{B} { B^{(k^\prime)}
}^\dagger(t^{\prime})\!\right\}$ and  $T_2(t-t^\prime)={\rm Tr}
_{B}\!\!\left\{\sum_{k,k^\prime=1}^{2}
{{B}^{(k)}}^\dagger(t){\rho}_{B}  {
B^{(k^\prime)}}(t^{\prime})\!\right\}$, respectively. In this
case, because the qubit's environments are totally uncorrelated,
we have $T_1(t-t^\prime)=T_2(t-t^\prime)=0$ if $k \neq k^\prime$.
We also suppose a simple case where the Coulomb coupling is null
$J=0$. Initially, the two DQDs are considered in a maximal quantum
correlated state given by $\rho_I(0)=|\Phi\rangle\langle\Phi|$,
where
$|{\Phi}\rangle=(|\!{\uparrow\downarrow}\rangle+|\!{\downarrow\uparrow}\rangle)/\sqrt{2}$.

\begin{figure}[t]
\begin{center}
\includegraphics[scale=.94]{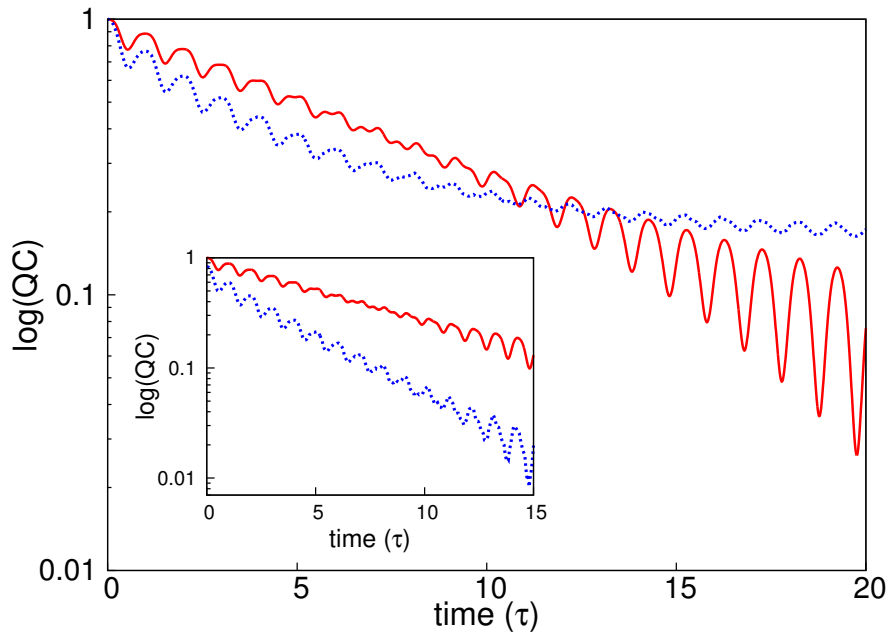}
\caption{(Color online) Logarithmic scale of EoF (solid curve) and
QD (dotted curve) as a function of time, considering $J=0$ and a
common bath with $T=0.1K$ and $\eta=1/600$. In the inset we plot the EoF curve and the QD curve minus the long time saturation value to extract only its exponential behavior.}\label{fig5}
\end{center}
\end{figure}

By examining Fig.~\ref{fig4}, we observe that for independent
environments the EoF and the QD exponentially tends to zero.
 On the other hand, for common environments, the QD has an exponential decay that tends
  to a finite value due to the specific form of the coupling between the system and the environment.
The fact that the environment acts correlated with both DQDs makes
a more robust QD. To investigate this robustness, in
Fig.~\ref{fig5} we compare the EoF and QD dynamics considering
only the common bath case and observe that QD is more resistant to
the environment's disturbance.
Moreover, the inset of Fig.~\ref{fig5} shows only the exponential
decay of EoF and the QD, \emph{i.e.}, the EoF curve and the QD curve minus the long time saturation value.
In this case, we notice that the exponential behavior of QD decays faster than EoF.

Based on our results, we conclude that the quantum correlation
dynamics of the two DQDs can be summarized as follows: the Coulomb
interaction introduces the correlations between both DQDs and the
environment acts in a way to preserve these
correlations for a longer time. Despite the fact that the
entanglement decays exponentially to zero, the QD is sustained
depending on the way the DQDs are coupled to the
environment.

\section{SUMMARY}
We have numerically solved the Redfield master equation to study
the quantum correlation dynamics of two DQDs at finite
temperatures including the interaction between the qubits and
detuning. We have verified that dissipation in
this system has an impact on the EoF which is stronger than on
the QD, thereby causing a shorter duration of the EoF. We have
shown that the QD is not completely destroyed through the
interaction with the environment and that it remains stable even
at finite temperatures. To explain these facts, we have explored
the characteristics of the bath, \emph{i.e.}, whether both qubits
are coupled simultaneously to a single bath or each qubit is
coupled to its own bath. The results have shown that even though
the coupling between the two-qubits is zero, the quantum
correlations survive for longer periods in the
case of a common bath. This behavior results from
the particular form of the interaction Hamiltonian and there are
systems for which the robustness of QD may not occur and each
different case must be analyzed carefully.

As a final comment, it should be desirable to study the same
effects in the context of the dynamics of a couple of Brownian
particles in a common bath which has recently been analyzed in
\cite{duarte} and where it has been shown that the reservoir
mediates an effective coupling between the particles. The
alternative (and more general) coupling to the environment used in
this problem could also be applied to the present model aiming at
a possible maintenance of the quantum correlations between the
qubits for a much longer time interval.

\section{Acknowledgements}
We wish to thank the partial financial support from the Conselho
Nacional de Desenvolvimento Cient{\'\i}fico e Tecnol{\'o}gico
(CNPq) and the Funda\c{c}{\~a}o de Amparo {\`a} Pesquisa no Estado
de S{\~a}o Paulo (FAPESP). AOC also acknowledges his participation
as a member of the Instituto Nacional de Ci{\^e}ncia e Tecnologia
em Informa\c{c}{\~a}o Qu{\^a}ntica (INCT-IQ).
We also would like to thank T. Werlang for reading the manuscript and for useful discussions.

\end{document}